\documentclass{emulateapj}
\def\kms{km\,s$^{-1}$\,}
\def\vs{$v_{S}$\,\,}
\def\etal{ et~al.\rm\,}

\def\dem71{DEM L 71\,}
\def\chan{{\it Chandra\,}}
\def\bn{$I_{B}/I_{N}$\,}
\def\tetp{($T_{e}/T_{p}$)$_{0}$\,}

\def\memp{$m_{e}/m_{p}$\,}

\tighten
\begin{document}
\submitted{Accepted November 9, 2006}

\title{A Physical Relationship Between Electron-Proton Temperature  
Equilibration and Mach
Number in Fast Collisionless Shocks  }

\author{Parviz Ghavamian\altaffilmark{1}, J. Martin Laming 
\altaffilmark{2} \& Cara E. Rakowski
\altaffilmark{2}  }

\altaffiltext{1}{Department of Physics and Astronomy, Johns Hopkins  
University, 3400 N. Charles
Street, Baltimore, MD, 21218; parviz@pha.jhu.edu}
\altaffiltext{2}{Naval Research Laboratory, Code 7674L, SW,  
Washington, DC; laming@nrl.navy.mil,
crakowski@ssd5.nrl.navy.mil}

\begin{abstract}

The analysis of Balmer-dominated optical spectra from non-radiative  
(adiabatic) SNRs
has shown that the ratio of the electron to proton temperature at the  
blast wave
is close to unity at \vs\,$\lesssim$400 \kms, but declines sharply  
down to the minimum value
of \memp\, dictated by the jump conditions at shock speeds exceeding  
2000 \kms.
We propose a physical model for the heating of electrons
and ions in non-cosmic ray dominated, strong shocks (\vs $> 400$ km s 
$^{-1}$)
wherein the electrons are heated
by lower hybrid waves immediately ahead of the shock front.  These  
waves arise naturally from
the cosmic ray pressure gradient upstream from the shock.  Our model  
predicts a nearly constant level
of electron heating over a wide range of shock speeds, producing a  
relationship
\tetp\,$\propto$\,$v_{S}^{-2}$ ($\propto\,M^{-2}$) that is fully  
consistent with the observations.

\end{abstract}

\section{INTRODUCTION}

The discovery of collisionless shock waves in the solar wind in the  
1960s ushered in
a new era in the physics of space plasmas (see Tidman \& Krall 1971,  
Sagdeev 1979 and
Kennel 1985 for reviews and references).  Due to the low density (n\,$ 
\lesssim$1 cm$^{-3}$) of
the interplanetary medium, the jump in hydrodynamical quantities is  
produced
not by Coulomb collisions but by collective plasma processes such as  
electromagnetic waves
and turbulence.  However, despite the availability of extensive in situ
observations of solar wind shocks and the expenditure of considerable  
effort in
theoretical modelling of these structures, a detailed understanding  
of the
processes at the shock transition responsible for partitioning the  
shock energy
between different charged particle species has been slow to emerge.   
The problem
is far more acute for interstellar shocks, where in situ measurements  
of the
the shock structure are unavailable and the very high Mach numbers ($ 
\sim$30$-$200)
make numerical simulations of these structures extremely difficult or  
impossible.

The optical emission generated by non-radiative supernova remnants  
(SNRs) in partially
neutral gas provides a valuable diagnostic tool for probing the  
heating processes in
collisionless shocks.
The optical spectra of these SNRs (which lose a negligible fraction  
of their energy
to radiation) are dominated by Balmer line emission,
produced by collisional excitation when neutral hydrogen is overrun  
by the blast
wave (Chevalier \& Raymond 1978, Bychkov \& Lebedev 1979).  Each  
emission line consists of two components:
(1) a narrow velocity component produced when cold, ambient H~I  
overrun by the shock is excited by electron
and proton collisions,
and (2) a broad velocity component produced when fast neutrals  
created by postshock charge exchange are
collisionally excited (Chevalier, Kirshner \& Raymond 1980).  An  
example of a Balmer-dominated shock
spectrum from the Galactic SNR RCW 86 is shown in Fig.~1.  The  
optical emission arises in a very thin
($\lesssim\,$10$^{16}$ cm) ionization zone, thin enough so that the  
protons transformed into hot neutrals have had
little time to equilibrate with electrons and other ions.  This makes  
the measured width of the broad Balmer
line directly proportional to the proton temperature set by  
collisionless heating at the shock front.  The broad-to-narrow flux
ratio, \bn\, on the other hand, is sensitive to both the degree of  
electron-proton temperature equilibration at the shock front
(i.e., \tetp) and the shock velocity, \vs\, (Chevalier, Kirshner \&  
Raymond 1980,
Smith \etal\, 1991).  The ratio also depends (though less  
sensitively) on the preshock neutral fraction.
The broad component width and \bn\, of
an observed Balmer-dominated shock can be modeled with numerical  
shock codes to simultaneously
estimate \vs and \tetp (Smith \etal\, 1991, Ghavamian 1999,
Ghavamian \etal\, 2001, Ghavamian \etal\, 2002).

In this paper we draw together observed values of \tetp and \vs measured
in five Balmer-dominated SNRs, including several previously  
unpublished measurements from
the SNR RCW 86. We then propose a physical model of electron heating  
by lower hybrid waves
in a cosmic-ray precursor that obeys the observed relationship  
between \tetp and \vs. We
conclude by exploring some consequences of this interpretation, the  
connection to
other observables and applicability to other collisionless shock  
situations.

\begin{figure}
\noindent \includegraphics[width=3.3in]{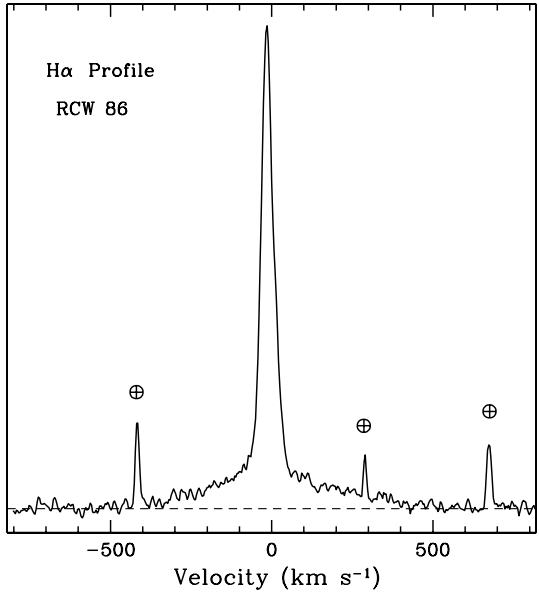}
\figcaption{ An example of the optical
spectrum of a Balmer-dominated shock, showing the broad and narrow
H$\alpha$ lines characteristic of non-radiative shocks in partially
neutral gas. This spectrum, originally presented by Sollerman
\etal\, (2003), was obtained from the southwestern rim of the
Galactic SNR RCW 86, with high enough spectral resolution ($\sim$10
\kms) to resolve the broad ($\sim$ 500 \kms\, FWHM)  and narrow
($\sim$30 \kms\, FWHM) H$\alpha$ lines.  The night sky OH lines
(indicated by the $\oplus$ symbol) have been left in to demonstrate
their relatively narrower widths compared to the H$\alpha$ lines.
The broad H$\alpha$ width and ratio of the broad to narrow H$\alpha$
flux for these types of shocks were used to produce the relationship
shown in Fig.~2.  }
\end{figure}

\section{Relationship Between Equilibration and Shock Speed}

In Fig.~2 we plot 11 measurements of \tetp\, and \vs, obtained
from longslit spectra of four Galactic remnants: the Cygnus Loop (one  
position from the northeast), RCW 86
(positions from the southwestern, northern and eastern limbs),  
Tycho's SNR ('Knot g' from the
eastern rim), SN 1006 (one position from the northwestern rim) and  
one remnant in the
Large Magellanic Cloud \dem71\, (five positions from the full  
circumference of the shell).
In the first four remnants, the shock
parameters have been estimated via longslit spectroscopy.  First, we  
used the broad component H$\alpha$ width
to constrain the range of shock speeds for each Balmer filament in  
the limits of minimal (\tetp\,=\,\memp)
and full (\tetp\,=\,1) equilibration.  Next, we fine tuned \tetp\,  
and \vs\, by using shock models to
match the observed broad to narrow flux ratio (Ghavamian \etal\,  
2001; 2002).

In the case of RCW 86 we have added two more data points, indicated  
by dashes in Fig.~2,
derived from previously unpublished optical spectra of Ghavamian (1999).
These were obtained from the eastern rim ($v_{FWHM}$(H$\alpha$)\,=\, 
640$\pm$35
\kms; \bn =1.0$\pm$0.2) and the northern rim ($v_{FWHM}$(H$\alpha$)\,=\,
325$\pm$10 \kms; \bn = 1.06$\pm$0.1) of RCW 86.
In the fifth SNR of our sample, \dem71 (Ghavamian \etal\, 2003),
we used the broad component H$\alpha$ width only to constrain the range
of shock speeds and the proton temperature $T_p$, due to
the anomalously low \bn\, values in this SNR (see below).  We then  
combined this information with
$T_e$ measured from \chan\, X-ray spectra of the blast wave to obtain  
\tetp\,
(Rakowski, Ghavamian \& Hughes 2003; Rakowski 2005).

Although the equilibrations
were obtained from different SNRs in a range of environments and  
distances, they show
a clear trend of decreasing \tetp with \vs.
The plot in Fig.~2 suggests that a transitional shock speed exists
around 400 \kms, below which collisionless processes promptly  
equilibrate the electrons and
protons at the shock front.  Above that speed \tetp\, rapidly  
declines, eventually reaching
values consistent with mass-proportional heating. This decline  
appears to follow \tetp\,$\propto\,v_S^{-2}$.
We show that this behavior is predicted by a physical model of  
electron heating,
where the electrons are heated in the cosmic ray precursor
to a level that is nearly independent of shock speed before acquiring  
the
mass-proportional increment at the shock front.  The protons, on the  
other hand,
receive only mass-proportional heating, resulting in the inverse  
squared dependence
of equilibration on shock speed.
Note that while our proposed model relies on the acceleration of  
cosmic rays and
the existence of a cosmic ray precursor to heat the electrons, it  
does not
require the energetics of the shock to be dominated by the cosmic rays.

In physical models the strength of the (assumed quasi-perpendicular)
shock is characterized not by the shock speed,
but rather the magnetosonic Mach number $M_S$ ($\equiv\,v_S / v_{MS} 
$, where
$v_{MS} \equiv (c_{S}^{2} + v_{A}^{2})^{1/2}$ is the magnetosonic speed,
$c_{S}$ is the sound speed ($=\sqrt{\frac{5}{3}\frac{P}{\rho}}$) and  
$v_{A}$ ($\equiv\,B/\sqrt{4\pi\,\rho_i}$)
the Alfv\'en speed of the preshock gas).  The preshock temperature,
ion density and magnetic field strength are not strongly constrained  
in the observed Balmer-dominated shocks.
In particular, $M_{S}$ is most sensitive to the choice of preshock  
magnetic field
due to the dependence of the Alfv\'en speed on $B^2$.  However,  
assuming standard
values for the warm neutral ISM - preshock temperature of 10,000~K,  
density of 1 cm$^{-3}$, magnetic field
strength of 3 $\mu$G and 50\% preshock ionization, the magnetosonic  
speed is then approximately
13 \kms.  The corresponding values of $M_S$ are marked at the top of  
Fig.~2.

\section{Lower Hybrid Wave Heating Model}

The constant electron heating with shock velocity (giving
\tetp\,$\propto v_S^{-2}$) suggests a process occurring in the
preshock medium rather than the shock front itself. Waves in the
preshock medium can be excited by shock reflected ions which gyrate
around the field lines before returning to the shock itself, or by a
cosmic ray precursor. Cargill \& Papadopoulos (1988) performed hybrid  
simulations
of the first possibility, wherein shock reflected ions generate  
Langmuir and ion-acoustic
waves ahead of the shock, which then heat electrons as they are  
damped. They predicted a temperature
ratio \tetp $\sim 0.2$ nearly independent of shock velocity, a result  
which clearly disagrees
with the observational data in Fig.~2. Laming (2001a,b) modeled the  
excitation of lower hybrid
waves by shock reflected ions, following the suggestion of McClements  
et al. (1997) that lower-hybrid
waves could stream away from the shock with a group velocity equal to  
the shock velocity. In this
way, waves stay in contact with the
shock for arbitrarily long periods of time and hence can grow to  
large intensities despite low
intrinsic growth rates. The electron heating predicted by such a model
(Vink \& Laming 2003) depends on the product of the maximum electron  
energy and the fraction
of electrons which are resonant with the waves that can be  
accelerated.  This gives
\tetp $\propto v_S^{-1}\times\exp\left(-M^2\right)$, where $M$ is the  
shock Mach number.

Here we explore another approach. Cosmic rays have been long known to  
generate
waves upstream of shocks; the generation of Alfv\'en waves is an  
intrinsic part of cosmic
ray acceleration models. Drury \& Falle (1986) showed that the  
negative gradient of cosmic pressure
with distance ahead of the shock can generate sound waves via the  
``Drury instability''. This is the
mechanism which smoothes out the hydrodynamical jump in cosmic ray  
modified shocks. Mikhailovskii (1992) gives a
corresponding expression for the excitation of magneto-acoustic  
waves. We argue that the high frequency
extension of such processes by the cosmic ray pressure gradient would
be the excitation of lower-hybrid waves.  The frequencies of these  
waves would lie between
the gyrofrequencies of the electrons and the protons, with electron  
heating occurring as the waves
damp along magnetic field lines.  Without calculating the
growth rate of the lower hybrid waves explicitly, we can estimate the  
magnitude of the electron
heating.  The parallel diffusion coefficient for electrons in lower  
hybrid wave turbulence is
(derived from Equations 10.83 and 10.93 of Melrose (1986); see also  
Begelman \& Chiueh 1988)
\begin{equation}
D_{\Vert\Vert}={1\over 4}\left(q\delta E\over m_e\right)^2
{\omega ^2\over k_{\perp}^2v_{\Vert}^2k_{\Vert}v_{\Vert}}={1\over 4} 
\left(q\delta E\over m_e\right)^2
{k_{\Vert}^2\over k_{\perp}^2}{1\over\omega}
\end{equation}
where $\omega =\sqrt{\Omega _e\Omega _p}$ is the lower hybrid wave  
frequency, the geometric mean of the
electron and proton cyclotron frequencies ($\Omega_{e,p} \equiv\,e  
B / m_{e,p} c$), and $k_{\Vert}$ and
$k_{\perp}$ are wavevectors parallel and perpendicular to the  
magnetic field.
Taking $\delta E=B\,\left(\frac{\Omega_{p}}{\omega}\right)^{1/3}\, 
\frac{\omega}{4 k_{\perp} c}$ (Karney 1978),
$k_{\Vert}^2/k_{\perp}^2=m_e/m_p$, and the time spent by an electron  
inside the cosmic ray
precursor $t\sim \kappa/v_S^2$ where $\kappa$ is the cosmic ray  
diffusion coefficient, we find
\begin{eqnarray}
{1\over 2}mv_e^2  =  {1\over 2}mD_{\Vert\Vert}t\nonumber \\
 = {m_e\over 128}\left(qB \over m_e\right)^2\left(\Omega _p\over 
\omega\right)^{2/3}{\omega ^2\over k_{\perp}^2c^2}{m_e\over m_p}{1 
\over\omega}{\kappa\over v_S^2}\nonumber\\
  =
{m_e\over 128}\Omega _e\left(\Omega _p\over\omega\right)^{5/3}{\omega  
^2\over k_{\perp}^2v_S^2}\,\kappa
\end{eqnarray}
If the lower hybrid wave group velocity $\partial\omega /\partial k_ 
{\perp} = v_S$, then
the phase velocity $\omega /k_{\perp} =2v_S$ when $k_{\Vert}^2/k_ 
{\perp}^2=m_e/m_p$ (Laming 2001a), and
the electron heating depends on a few constants times the product $ 
\Omega _e\kappa$.
Thus for a Bohm-like cosmic
ray diffusion coefficient $\kappa\propto 1/B$ and independent of \vs,  
we have an electron heating
process independent of both the shock velocity and the upstream  
magnetic field,
and \tetp\,$\propto\, v_S^{-2}$ as required.
This estimate neglects the fact that only electrons with velocity  
greater than $v_e\sim\omega /k_{\Vert}$
may be heated by the waves. However, for the likely depth of a cosmic  
ray precursor (estimated below),
the electrons will have sufficient time to at least partially  
collisionally equilibrate amongst themselves
before crossing
the shock, allowing a much larger fraction of preshock electrons to  
interact with the waves.
The time spent by an electron inside a reflected ion precursor,
as opposed to a cosmic ray precursor, would be $t\sim d/v_S\sim 1/ 
\Omega _p$, where the precursor depth $d$
is approximately the gyroradius of a cosmic ray proton.
Substituting this relation into equation 2 gives \tetp independent of  
\vs,
similar to the behavior predicted by Cargill \& Papadopoulos (1988),
although these authors consider different waves. In this case the  
shallowness
of the reflected ion precursor prevents the collisional redistribution
of energy amongst the electrons.

\begin{figure}
\noindent \includegraphics[width=3.4in]{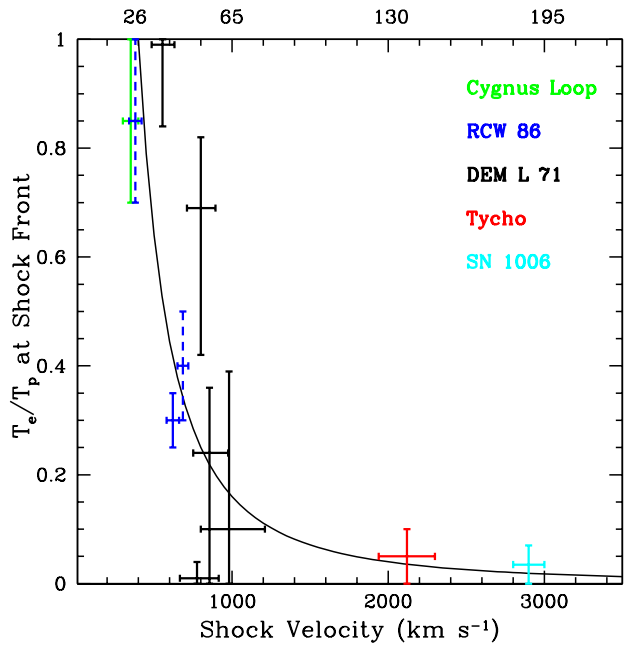}
\figcaption{The electron to proton temperature ratio at the shock front
as a function of shock velocity for 5 Balmer-dominated SNRs.
Magnetosonic Mach numbers
($M_S$) appropriate for typical ISM conditions are indicated along
the top axis.
The data shown here were measured from Balmer-dominated
shocks in the Cygnus Loop, RCW 86, Tycho's SNR (Ghavamian \etal\,
2001), SN 1006 (Ghavamian \etal\, 2002) and DEM L71 (Rakowski \etal\
2003, Rakowski 2005).
The dashed error bars for RCW 86 mark previously unpublished results.
Below 400 \kms\, ($M_S\,\approx\,$30) the data are consistent with
\tetp\,=\,1.
The prediction of the proposed lower hybrid wave heating mechanism in
the cosmic-ray precursor,
\tetp\,$\propto$\,$V_{S}^{-2}$ ($\propto\,M_S^{-2}$), is shown for \vs
\,$>$\,400 \kms.  }
\end{figure}

If a constant kinetic energy is imparted to electrons by lower hybrid  
waves, then for shock speeds
of 400 \kms\, and above (c.f. Fig.~2) the observations require  
$m_ev_e^2/2\,\sim\,4\times10^{-10}$ ergs ($\sim$ 0.3 keV)
immediately behind the shock front.  In that case, we infer $\Omega _e 
\kappa\simeq 7\times
10^{21}$ cm$^{2}$s$^{-2}$, and $\kappa\sim 4\times 10^{14}/B$ cm$^2$s 
$^{-1}$. For typical magnetic
fields $B\sim$3 $\mu G$, $\kappa$ is several orders of magnitude  
below that inferred for the undisturbed
interstellar medium, but entirely consistent with values estimated  
for the solar wind or interplanetary
medium. We emphasize that this estimate is a lower limit on $\kappa$.  
This estimate neglects the
electron-electron collisional equilibration necessary to bring more  
electrons into resonance with
the turbulence. Furthermore, lower hybrid waves can only be excited  
in the cosmic ray
precursor if the cosmic ray pressure gradient is sufficiently strong.  
The required gradient scale lengths
are $1/L > 8c_S/3\kappa$ for sound waves (Drury \& Falle 1986), and
$1/L > v_A/cr_g\sim v_A/\kappa$ for magneto-acoustic waves  
(Mikhailovskii 1992), where $r_g$ is the gyroradius of
a cosmic ray proton.   An upper limit on $\kappa$ will come from the  
requirement that the neutral H
survive against electron impact ionization in the precursor to reach  
the shock front. Numerically,
$\kappa \lesssim 10^{24}\left(v_{S}/1000 {\rm ~km~s}^{-1}\right)^2/n_e 
$ cm$^2$s$^{-1}$.
Further constraints on the cosmic ray diffusion coefficient and  
pressure will be obtained from
calculation of the lower hybrid wave growth rate required to sustain  
the electron heating.  This
growth rate must be significantly larger than the charge exchange  
frequency of the partially
neutral gas.  Under typical ISM conditions $\sqrt{\Omega_{e} \Omega_ 
{p}}  \sim$ 1 Hz
and $n_{H^{0}}\langle\sigma_{cx} v\rangle \sim\,$
10$^{-8}$ Hz, so the condition for electron heating is nearly always  
satisfied.

The electron heating by lower-hybrid waves is inherently anisotropic.  
If some of this anisotropy
survives the electron-electron collision equilibration, a  
polarization signal may be present in the
narrow component of H$\alpha$ (see e.g. Laming 1990), which might  
provide more insight into the electron
heating process.

\section{The Width of the H$\alpha$ Narrow Component}

The same cosmic ray precursor should also generate lower frequency  
waves, which for a quasi-perpendicular
shock will include magneto-acoustic waves. Below the ion-neutral charge
exchange frequency these waves are not effectively damped (Drury \etal 
\, 1996),
and may reveal themselves via broadening of the narrow H$\alpha$  
component (Smith, Raymond, \& Laming 1994).
Those authors suggested that waves in the cosmic ray precursor  
actually heat the preshock gas to 30,000 - 40,000 K and that
the observed linewidth is thermal in nature. An upper limit to the  
cosmic ray diffusion coefficient
of $\sim 10^{24}/n_e$ cm$^2$ s$^{-1}$ then results from the  
constraint that sufficient neutral H survive
passage through the precursor to encounter the shock. However no  
precise heating mechanism was specified,
and we now suggest that the large width of the narrow component H$ 
\alpha$ line in the observed SNRs
is not of thermal origin, but rather due to the motion of protons in  
the lowest frequency waves of
the magneto-acoustic spectrum.  These waves lie below the charge  
exchange frequency, allowing a coherent
oscillation of preshock neutrals and preshock protons.  In this case,  
$\delta v\simeq v_A\delta B/B$ for
Alfv\'en waves ($v_A$ should be replaced by $v_{MS}$ for magneto- 
acoustic waves).
The Alfv\'en speed upstream from the shock is in the range 1-10 km s$^ 
{-1}$, so an observed broadening
$\delta v/v_A >1$ also implies $\delta B/B >1$.
Such a magnetic field amplification has already been proposed in the  
context
of cosmic ray acceleration in shocks (Lucek \& Bell 2000, Bell \&  
Lucek 2001).
A successful model of precursor ions will need to reproduce the  
existing measurements
of narrow component H$\alpha$ line widths (Smith, Raymond \& Laming  
1994; Hester \etal\, 1994; Sollerman
\etal\, 2003) which indicate that they remain relatively constant  
(30-50 \kms) over a wide range in
shock speed (180-3000 \kms). We defer this topic to future work.

\section{Discussion and Summary}

The plot of \tetp in Fig. 2 is qualitatively similar to results of a  
survey of electron heating at solar wind
shocks (Schwartz et al. 1988). The main difference is that $T_e\sim  
T_i$ is obtained for Alfv\'enic
Mach numbers $\le 5$ in the solar wind, whereas in Fig. 2 this occurs  
for magnetosonic 
Mach numbers around 20-30, for assumed preshock magnetic fields of 3$ 
\mu$G. A preshock
magnetic field amplification in SNRs, of a similar amount required to  
reproduce the width of
the H$\alpha$ narrow component ($\delta B/B\,\sim\,$ 5-10), would  
shift the \tetp ratios in Fig. 2 onto Mach
numbers similar to those in the solar wind.  This may support our  
interpretation.
Collisionless shocks play a dominant role in heating the  
intergalactic medium (IGM) during large scale structure formation
(Ryu \etal\, 2003; Yoshida \etal\, 2005; Kang \etal\, 2005).
If the preshock magnetic field is amplified as described above,
then the relationship \tetp\,$\propto\,M_{S}^{-2}$ may also be  
applicable to modeling the ionization and
temperature structure of the IGM.

The maximum energy attainable by cosmic rays in Balmer-dominated
shocks is limited by ion-neutral damping to $\sim$0.1-1 TeV (Drury  
\etal\, 1996).
It is possible that the shocks with stronger cosmic ray acceleration  
generate sufficient
electron heating in their precursors that no neutrals survive to  
cross the shock. The anomalously
low \bn ratio observed in DEM L71 (Ghavamian \etal\, 2003) may  
indicate that this is an
intermediate case. Electron heating in a
shock precursor will produce added narrow component H$\alpha$.   
However, since the broad component can
only arise by charge exchange with shocked protons, the neutral H  
must penetrate either the shock
or the reflected ion precursor. We emphasize that of the points  
plotted in Fig. 2 from H$\alpha$ \bn
measurements, the electron temperatures from RCW 86 (Vink et al.  
2006), the Cygnus Loop
(Levenson et al. 2002), SN 1006 (Laming et al. 1996, Vink et al.  
2003) and Tycho (Hwang et al. 2002;
Warren \etal\, 2005) have been independently corroborated by  
measurements in other wavebands.

Several lingering questions remain from Fig.~2:
what mechanism causes the prompt electron-proton equilibration below  
400 \kms, and why does
the decline in \tetp\, begin at that shock speed?  Even more  
importantly, does the inverse square
relationship between equilibration and shock speed/Mach number also  
hold for collisionless shocks
in fully ionized gas? These issues will be addressed in future
work.

\acknowledgements
P.G. would like to thank R. Cen and K. Sembach for helpful  
discussions and acknowledges support from
NASA Contract NAS8-03060.  JML and CER acknowledge support from NASA  
Contract
NNH06AD66I (LTSA Program) and basic research funds of the Office of  
Naval Research.

\end{document}